\documentclass[prb, reprint, superscriptaddress, longbibliography]{revtex4-2}
\usepackage{amsmath}
\usepackage{amssymb}
\usepackage{graphicx}
\usepackage[caption=false, position=top, singlelinecheck=off, justification=raggedright]{subfig}
\usepackage{multirow}
\usepackage{color}
\usepackage[dvipsnames]{xcolor}
\usepackage{ulem}
\usepackage{pst-node}
\usepackage{appendix}
\usepackage[unicode]{hyperref}
\hypersetup{
	unicode=true,         	
	colorlinks=true,      	
	linkcolor=blue,		   	
	citecolor=blue,       	
	urlcolor=blue		   	
}

\begin{document}

\title{Optimal parametric control of transport across a Josephson junction}

\author{Hannah Kleine-Pollmann}
\affiliation{Zentrum f\"ur Optische Quantentechnologien and Institut f\"ur Quantenphysik, 
	Universit\"at Hamburg, 22761 Hamburg, Germany}

\author{Guido Homann}
\affiliation{Zentrum f\"ur Optische Quantentechnologien and Institut f\"ur Quantenphysik, 
Universit\"at Hamburg, 22761 Hamburg, Germany}

\author{Ludwig Mathey}
\affiliation{Zentrum f\"ur Optische Quantentechnologien and Institut f\"ur Quantenphysik, 
Universit\"at Hamburg, 22761 Hamburg, Germany}
\affiliation{The Hamburg Centre for Ultrafast Imaging, Luruper Chaussee 149, 22761 Hamburg, Germany}

\date{\today}
\begin{abstract}
We present optimal control strategies for the DC transport across a Josephson junction. Specifically, we consider a junction in which the Josephson coupling is driven parametrically, with either a bichromatic or a trichromatic driving protocol, and optimize the prefactor of the 1/$\omega$ divergence of the imaginary part of the conductivity. We demonstrate that for an optimal bichromatic protocol an enhancement of 70 can be reached, and for an optimal trichromatic protocol an enhancement of 135. This is motivated by pump-probe experiments that have demonstrated light-enhanced superconductivity along the c-axis of underdoped YBCO, where the junction serves as a minimal model for the c-axis coupling of superconducting layers. Therefore, the significant enhancement of superconductivity that we show for multi-frequency protocols demonstrates that the advancement of pump-probe technology towards these strategies is highly desirable.
\end{abstract}
\maketitle

\section{Introduction}
Pump-probe experiments on materials have established remarkable phenomena, such as light-enhanced and light-induced superconductivity in high-temperature superconductors such as YBCO, and other materials~\cite{Hu:2014aa,Fausti:2011}. These phenomena consist of either enhancing the superconducting response of a material that is a superconductor initially, or achieving a superconducting response of a material that is initially above the critical temperature.~\cite{Foerst:2014, Foerst:2015, Kaiser:2014,Nicoletti:16,Averitt:2019,Shimano:2023,Shimano:2023(2)} The observation of these phenomena suggests that dynamical control of materials via light is possible. This poses the question of how and how much a material property can be controlled in a desirable manner.
A proposed mechanism for light-enhanced superconductivity in cuprates is parametric enhancement, exemplified for a Josephson junction in~\cite{Okamoto-2016}. Here, the Josephson coupling is modulated parametrically, resulting in an enhancement of the DC transport across the junction, which can also be expressed as an enhancement of the effective Josephson coupling. Specifically, this mechanism shows a maximal increase of the effective Josephson interlayer coupling of a factor of 2 - 3 in a steady state, when the driving frequency is blue-detuned, compared to the plasma frequency of the Josephson junction.
In the transient regime, rather than the steady state, the enhancement of the effective Josephson coupling exceeds the bare coupling further~\cite{Okamoto-2017}.
This mechanism also has an influence on Meissner screening~\cite{Guido:2022:Meissner}, and can be demonstrated in Josephson-coupled condensates ~\cite{Beilei:2021}. Furthermore, related phenomena in the context of non-linear coupling between the electric field and the Higgs mode have been discussed~\cite{Guido:2021, Guido:2020, Ruebenhaus:2023}, as well as photo-induced eta-pairing~\cite{Kaiser:2019}.

\begin{figure}[h!]
	\centering
	\includegraphics[scale=1]{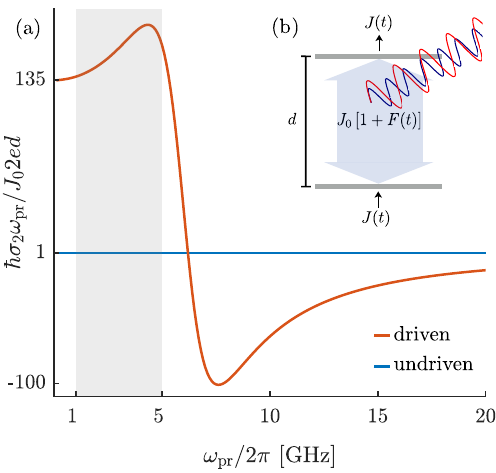}
    \caption{\textbf{(a)} Imaginary part of the conductivity $\sigma_2 (\omega_{\mathrm{pr}})$ as a function of the probing frequency $\omega_{\mathrm{pr}}$ in the steady state.
    The undriven case (blue) is compared to the case with an optimal trichromatic driving protocol (red). The gray area depicts the interval in which the conductivity is evaluated numerically to estimate the effective Josephson coupling $J_{\mathrm{eff}}/J_0$. For this optimal protocol, the DC transport across the junction is enhanced by a factor of 135.
    \textbf{(b)} 
    Schematic representation of a parametrically driven Josephson junction with the thickness $d$ and the Josephson coupling $J_0$.
    The driving protocol $F(t)$ contains several frequency components as indicated by the wavy lines.
    To measure the  conductivity, a probe current $J(t)$ is applied.
    } 
    \label{fig:fig1}
\end{figure}
In this paper, we present optimal parametric driving protocols of a Josephson junction~\cite{Josephson:1962}. With this, we address the question of how much the transport across a Josephson junction can be enhanced via optimal control.
We optimize the DC transport across the junction, based on, the low-frequency limit of the imaginary part of the conductivity.
The Josephson junction serves as a minimal model for the Josephson plasma mode dynamics of a cuprate superconductor.
Therefore, the transport across the junction is a minimal model for c-axis transport in a cuprate. Thus, we motivate optimal driving protocols of light-driven cuprates to maximally enhance the c-axis transport.

The magnitude of  the low-frequency limit of the imaginary part of the conductivity is quantified by the effective Josephson coupling, which is proportional to the prefactor of the 1/$\omega_{\mathrm{pr}}$ divergence of the imaginary part of the conductivity. Here, $\omega_{\mathrm{pr}}$ is the probing frequency.
We drive the Junction parametrically and probe the junction to read out the low frequency conductivity response in order to derive an effective Josephson coupling.
We utilize a Metropolis algorithm~\cite{Metropolis:1953, Ashton:2022} with the objective to enhance the effective Josephson coupling, in the steady state.
We optimize two classes of driving protocols.
First, we consider a bichromatic driving protocol and show that driving at the second harmonic leads to an enhancement of the effective Josephson coupling of around 70, compared to the undriven case.
Second, we consider a trichromatic driving protocol and find an enhancement of the effective Josephson coupling of around 135, compared to the undriven case.
Interestingly, this optimal enhancement is achieved for a set of frequencies, in which the third frequency is not at a simple integer ratio to one of the other frequencies, while the first and second frequency are in a 1:2 ratio. Furthermore, we find an enhancement of the effective Josephson coupling of around 50 for a wide range of frequencies.
These results suggest that multi-frequency driving protocols are a desirable method to optimally enhance the optical conductivity in high-$T_c$ superconductors, in the steady state.

This paper is organized as follows.
In Sec. II we introduce our method to estimate the effective Josephson coupling.
In Sec. III we introduce the numerical method to optimize the driving protocol.
In Sec. IV, we show the enhancement of the effective Josephson coupling for optimal bi- and trichromatic driving protocols.
In Sec. V we show the conductivity for optimal bi- and trichromatic driving protocols, in comparison to the undriven case.
In Sec. VI, we conclude our findings.

\section{Conductivity of a Josephson junction}
We consider a single Josephson junction~\cite{Beilei:2016} with thickness $d$, Josephson coupling $J_0$, and dielectric constant $\epsilon$.
In the absence of driving, the equation of motion for the phase $\varphi$ of the junction is given by the RCSJ model~\cite{RCSJ-Model}
\begin{equation}
    \Ddot{\varphi} + \gamma \Dot{\varphi} + \omega_{\mathrm{pl}}^2 \sin \varphi= \frac{\omega_{\mathrm{pl}}^2 J}{J_0}, \label{Eq:DGL}
\end{equation}
where $\omega_{\mathrm{pl}}= \sqrt{2ed J_0/\epsilon \epsilon_0 \hbar}$ is the plasma frequency of the junction and $\gamma$ is a damping constant.
Throughout this manuscript we use a plasma frequency of $\omega_{\mathrm{pl}}/2\pi=1\text{THz}$.
We probe the junction by applying a monochromatic current
\begin{equation}
    J= J_{\mathrm{pr}} \cos(\omega_{\mathrm{pr}} t),
\end{equation}
where $J_{\mathrm{pr}}$ is a probing current that is small compared to the Josephson coupling  $J_{\mathrm{pr}}/J_0 \ll 1$ and $\omega_{\mathrm{pr}}$ is the probing frequency.
Note that we focus on small probing frequencies $\omega_{\mathrm{pr}} < \omega_{\mathrm{pl}}$ in the following.
In frequency space, we compute the linear conductivity
\begin{equation}
    \sigma(\omega_\mathrm{pr})=\frac{J(\omega_{\mathrm{pr}}) d}{V(\omega_{\mathrm{pr}})},
    \label{eq:sig}
\end{equation}
where $V(t)=(\hbar/2e)\dot{\varphi}(t)$ is the junction voltage in the time domain, and $J = J(\omega_{\mathrm{pr}})$.
As we show in Appendix~\ref{App:Analytical}, the numerical result is in excellent agreement with the analytical solution for the equilibrium case
\begin{equation}
    \sigma(\omega_{\mathrm{pr}})= \epsilon \epsilon_0 \left[\gamma + i \left(\frac{\omega_{\mathrm{pl}}^2}{\omega_{\mathrm{pr}}} - \omega \right) \right] .
\end{equation}
Next, we add parametric driving to the dynamics
\begin{equation}
    \Ddot{\varphi} + \gamma \Dot{\varphi} + \omega_{\mathrm{pl}}^2 [1+F(t)] \sin \varphi= \frac{\omega_{\mathrm{pl}}^2 J}{J_0}, \label{eq:DGLdrving}
\end{equation}
where $F(t)$ is a driving protocol. A schematic representation of the driven system is shown in Fig.~\ref{fig:fig1} (b). 
We define the effective Josephson coupling as
\begin{equation}
    J_{\mathrm{eff}}=\frac{\hbar}{2ed}\left[\sigma_2(\omega_{\mathrm{pr}})\omega_{\mathrm{pr}}\right]_{\omega_{\mathrm{pr}} \rightarrow 0} .
    \label{eq:jeff}
\end{equation}
Note that $J_{\mathrm{eff}}=J_0$ in equilibrium. 
Numerically, we proceed as follows.
Once the system has reached a steady state, we estimate $\sigma_2(\omega_\text{pr})$ as the imaginary part of $\sigma(\omega_\text{pr})$ according to Eq.~(\ref{eq:sig}) where we record the dynamics of the phase $\varphi (t)$ over a time interval of $25000 \times T$, with $T= 2\pi/\omega_{\mathrm{pl}}$. 
Next, we evaluate the effective Josephson coupling $J_{\mathrm{eff}}$ from a linear fit of $\sigma_2(\omega_{\mathrm{pr}}) \omega_{\mathrm{pr}}$ in the range of $\omega_{\mathrm{pr}} \in [0.001 \omega_{\mathrm{pl}},~0.005 \omega_{\mathrm{pl}}]$ as shown as the gray shaded area in Fig.~\ref{fig:fig1}(a).
Because we estimate the effective Josephson coupling based on a finite number of samples we obtain an error $\epsilon_J$ as the standard deviation in units of $J_{\mathrm{eff}}$. This error is used below to exclude protocols for which $J_{\mathrm{eff}}$ cannot be determined with sufficient clarity.

\section{Monte-Carlo optimization}
We consider a parameterized driving protocol $F(t) = F_q(t)$, where $q=\{q_k\}$ is the set of driving parameters $q_k$ that parametrize the driving protocol.
We utilize the Metropolis algorithm~\cite{Metropolis:1953} to optimize the driving parameters $q$ with the objective to enhance the effective Josephson coupling $J_{\mathrm{eff}}/J_0$ according to Eq.~(\ref{eq:jeff}).

We initialize the driving parameters $q$ and calculate $J_{\mathrm{eff}}(q)$.
For perfect convergence of the algorithm, the choice of the initial driving parameters is arbitrary.
However, we find that specific initialization strategies, depending on the specific driving protocol $F_q(t)$, have a strong impact on the optimization  process, as discussed in the Appendix~\ref{Initial}.
Starting from an initial choice of the parameters $q_k$, the Metropolis algorithm generates a Markov chain of parameters, to search for an optimal protocol.
The update step of the algorithm generates a set of parameters $p_k$ out of the parameters $q_k$ via 
\begin{equation}
      p_k=q_k+x_k
      \label{eq:qt}
\end{equation}
where $x_{k}\sim U[-r_k,r_k]$ is uniformly sampled and $r_k$ is a hyperparameter which determines the maximum change of the $k$-th parameter. We define the set of parameters of maximum change as r = $\{r_k\}$.
We exclude a parameter set $p$ if $F_p(t)<-1$ for any time $t$. This ensures positive Josephson coupling, as is implied by the form of the driving term in Eq.~(\ref{eq:DGLdrving}), and $\omega_{\mathrm{pl}}^2 \sim J_0$. Depending on the specific physical implementation of the Josephson junction, this condition could be modified to capture the physically realizable range of $J_0$ or $\omega_{\mathrm{pl}}$. We also exclude regions where the driving protocol leads to Floquet parametric instabilities~\cite{floquet1,floquet2}, as further discussed in the Appendix~\ref{A2. Floquet}.
Next, we calculate the effective Josephson coupling $J_{\mathrm{eff}}(p)$, corresponding to the trial driving parameters $p$, based on Eq.~(\ref{eq:jeff}).
If we obtain an error of $\epsilon_J > 0.1$, we sample a new set of trial driving parameters $p$ according to Eq.~(\ref{eq:qt}).
Otherwise, we compute the probability $P(q, p)$ that the trial driving parameters $p$ are accepted over the current driving parameters $q$ as
\begin{equation}
    P(q, p) = \min\left(1,\mathrm{exp}\left(-\frac{J_{\mathrm{eff}}(q)-J_{\mathrm{eff}}(p)}{J_0 \alpha}\right) \right),
\end{equation}
where $\alpha$ is a constant hyperparameter.
Throughout this study we use a value of $\alpha=0.3$. We note that in the context of sampling a thermal ensemble,  the hyperparameter $\alpha$ is related to the temperature of the ensemble.
Next, we uniformly sample a number $z\sim U[0,1]$ and update the current driving parameters as $q \rightarrow p$ if $P(q, p) \geq z$.
We repeat this procedure for 1500 iterations.
We note that the possibility to accept $p$ as a new set of driving parameters even if $J(p) < J(q)$, reduces the probability to converge into local minima.

We consider two driving protocols $F(t)$.
First, we consider a bichromatic driving protocol
\begin{equation}
    F_B(t)=A_1\cos(\omega_1t) + A_2\cos(\omega_2t+\theta),
\end{equation}
which contains cosine modes of frequencies $\omega_1$ and $\omega_2$ with the amplitudes $A_1$ and $A_2$ and a phase shift $\theta$.
The set of parameters of maximum change is $r = \{0.05, 0.05, 0.05, 0.05, 2\pi \times 0.05\}$. 
Second, a trichromatic driving protocol
\begin{equation}
    F_T(t)=A_1\cos(\omega_1 t) + A_2\cos(\omega_2 t +\theta_1) + A_3\cos(\omega_3 t + \theta_2),
\end{equation}
which contains cosine modes with the frequencies $\omega_1$, $\omega_2$ and $\omega_3$ ,and the amplitudes $A_1$, $A_2$ and $A_3$ and the phase shifts $\theta_1$ and $\theta_2$. Here, the set of driving parameters is
$q = \{\omega_1, \omega_3, A_1, A_2, A_3, \theta_1, \theta_2\}$. We fix the frequency $\omega_2$ to $\omega_2 = 2\times \omega_1$, i.e. the second harmonic, as we discuss below. The set of parameters of maximum change is $r = \{0.05, 0.05, 0.05, 0.05, 0.05, 2\pi \times 0.05,2\pi \times 0.05\}$. 

\section{Results}
\subsection{Bichromatic driving protocol}

\begin{figure}[h!]
	\includegraphics[scale=1]{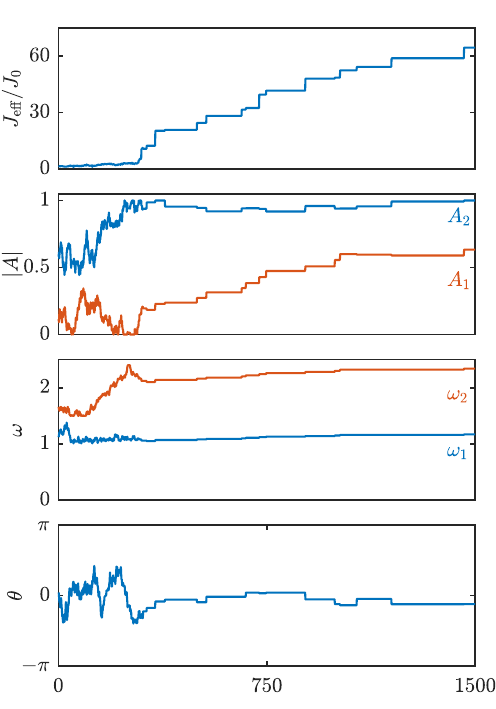}
    \caption{Optimization of the driving parameter over 1500 optimization iterations and the corresponding effective Josephson coupling $J_{\mathrm{eff}}/J_0$ for the bichromatic driving protocol $F_B(t)$. We observe a strong enhancement of the effective Josephson coupling $J_{\mathrm{eff}}/J_0$.}
    \label{fig:fig2}
\end{figure}

\begin{figure}[h!]
    \centering
    \includegraphics[scale=1]{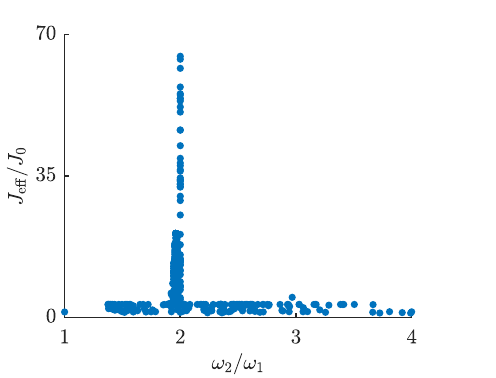}\llap{\parbox[b]{16cm}{\textcolor{black}{(a)}\\\rule{0ex}{6.4cm}}} \hspace{0.05cm} \includegraphics[scale=1]{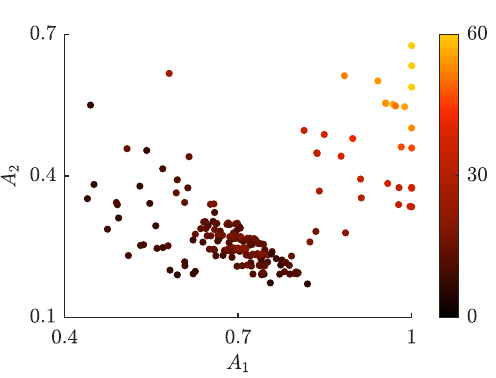}\llap{\parbox[b]{16cm}{\textcolor{black}{(b)}\\\rule{0ex}{6.4cm}}}
	\caption{Optimization of the bichromatic driving protocol $F_B(t)$. (a) Maximal effective Josephson coupling $J_{\mathrm{eff}}/J_0$ as a function of the driving frequencies $\omega_2/\omega_1$ for 600 optimization trajectories. The driving frequency $\omega_1$ is usually within the blue detuned regime, around $\omega_1/\omega_{\mathrm{pl}}\approx 1.1$. We obtain a strong enhancement at $\omega_2/\omega_1\approx 2$. 
    (b) Maximal effective Josephson coupling $J_{\mathrm{eff}}/J_0$ as a function of the driving amplitudes $A_1$ and $A_2$ for 600 optimization trajectories where we only depict trajectories having an effective Josephson coupling $J_{\mathrm{eff}}/J_0>10$.
    We obtain a strong enhancement for $A_1=1$.}
    \label{fig:fig3}
\end{figure}

To determine the optimal bichromatic driving protocol $F_B(t)$, we generate 600 optimization trajectories for various initial driving parameters $q_0$ (Tab.~\ref{tab:bichrom}).
In Fig. \ref{fig:fig2}, we show an example for an optimization trajectory.
We find that the effective Josephson coupling $J_\mathrm{eff}$ converges to significantly increased values up to $J_\mathrm{eff}/J_0 \approx 70$, compared to the Josephson coupling $J_0$. At the beginning of the optimization, the driving parameters $q$ change frequently.
As soon as a significant increase of the effective Josephson coupling $J_\mathrm{eff}/J_0$ is achieved, the driving parameters $q$ converge.
In almost all cases, we obtain a blue-detuned driving frequency $\omega_1$ with respect to the plasma frequency of around $\omega_1/\omega_\mathrm{pl}\approx 1.1$.
The second driving frequency converges close to  $\omega_2/\omega_1 \approx 2$, i.e. the second harmonic,  as we illustrate in Fig.~\ref{fig:fig3}(a).
 
In Fig.~\ref{fig:fig3}(b) we show the subspace of the driving amplitudes $A_1, A_2$ and the corresponding effective Josephson coupling $J_{\mathrm{eff}}/J_0$.
A driving amplitude of $A_1\approx 1$ and driving amplitude of $A_2 \approx 0.7$ leads to a significant increase of the effective Josephson coupling $J_{\mathrm{eff}}/J_0$ of around $70$.
Preferably, the phase difference $\theta$ is close to 0 for a maximized effective Josephson coupling $J_\mathrm{eff}/J_0$, which has been shown in previous studies~\cite{Okamoto-2017}. We obtain that a maximized amplitude $|F_B(t)|$ from the driving protocol $F_B(t)$ is preferable.
We observe  that we deviate from a phase difference of $\theta=0$ in order to avoid $F_B(t)<-1$ for all times, which would lead to a negative Josephson coupling, in favor of the maximized $|F_B(t)|$, see also Fig~\ref{fig:driving vergleich}.

Overall, we find a remarkable increase in the effective Josephson coupling $J_{\mathrm{eff}}/J_0$ of 71.69 with the driving parameters $A_1=1, \ A_2=0.6334  , \ \omega_1 / 2\pi=1.17 \ \mathrm{THz} , \ \omega_2 / 2\pi= 2.34 \ \mathrm{THz}$ and $\theta=-0.3812$.
\subsection{Trichromatic driving protocol}
Next, we discuss the case of the trichromatic driving protocol $F_T(t)$.
\begin{figure}[h!]
	\centering
	\includegraphics[scale=1]{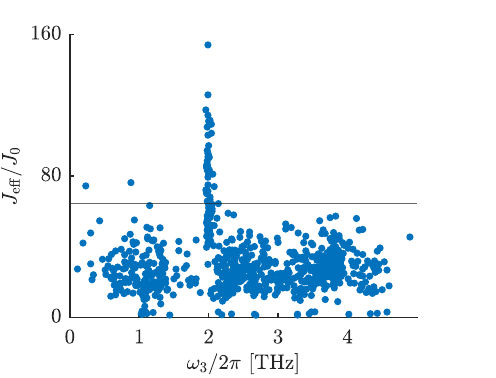}\llap{\parbox[b]{16cm}{\textcolor{black}{(a)}\\\rule{0ex}{6.4cm}}} \hspace{0.05cm} \includegraphics[scale=1]{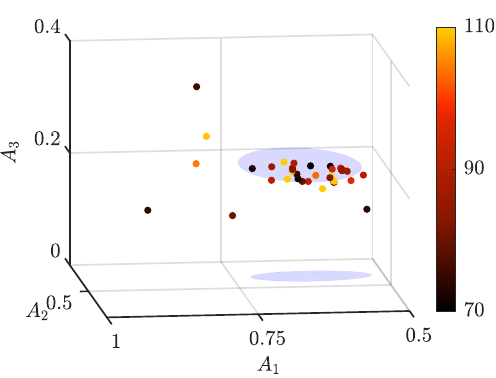}\llap{\parbox[b]{16cm}{\textcolor{black}{(b)}\\\rule{0ex}{6.4cm}}}
    \caption{Optimization of the trichromatic driving protocol $F_T(t)$. (a) Maximal effective Josephson coupling $J_{\mathrm{eff}}/J_0$ as a function of the driving frequency $\omega_3/2\pi$ [THz] for 950 optimization trajectories. The driving frequency $\omega_1$ is usually within the blue detuned regime, i.e. $\omega_1/\omega_{\mathrm{pl}}\approx 1.1$. We obtain a strong enhancement at $\omega_3\approx 2$. 
    The black line indicates the highest effective Josephson coupling $J_{\mathrm{eff}}/J_0$ ($\approx 70$) which we find with a bichromatic driving protocol $F_B(t)$.
    (b) Maximal effective Josephson coupling $J_{\mathrm{eff}}/J_0$ as a function of the driving amplitudes $A_1,A_2$ and $A_3$ for 950 optimization trajectories where we only depict trajectories having an effective Josephson coupling $J_{\mathrm{eff}}/J_0>70$. The shadowed elliptic areas are the projection on the $A_1,A_2$ and $A_1,A_3$ plane.}
    \label{fig:fig4}
\end{figure}

We set the driving frequency $\omega_2$ to the second harmonic of $\omega_1$ motivated by the optimal bichromatic driving protocol obtained above.
We generate 950 optimization trajectories for various initial driving parameters $q_0$ (Tab.~\ref{tab:trichrom}) in order to optimize the driving parameters $q$ of the trichromatic driving protocol $F_T(t)$.
In Fig.~\ref{fig:fig4}(a) we show the maximal effective Josephson coupling $J_{\mathrm{eff}}/J_0$ as a function of the driving frequency $\omega_3$ for all optimization trajectories. 
We find a significant increase of the effective Josephson coupling $J_{\mathrm{eff}}/J_0 \approx 155$ for driving frequencies $\omega_1\approx 1.1 \omega_{\mathrm{pl}}$ and $\omega_3 \approx 2 \omega_{\mathrm{pl}}$.
A driving amplitude of $A_1\sim[0.5-0.75] $ and a driving amplitude of $A_2 \sim[0.5-0.2]$ with a slightly smaller $A_3$ constitutes the optimal regime, as shown in Fig.~\ref{fig:fig4}(b).
Note that in contrast to the bichromatic driving protocol $F_B(t)$, the initial driving parameter $q_0$, especially $\omega_3$, are crucial for the convergence of the algorithm, which we further analyze in the Appendix~\ref{Initial}.

Overall we obtain a remarkable increase in the effective Josephson coupling $J_{\mathrm{eff}}/J_0$ of 155.
The corresponding optimal driving parameters are $A_1=0.6514, \ A_2= 0.2991, \ A_3= 0.1873, \ \omega_1/2\pi = 1.06 \ \mathrm{THz}, \ \omega_3/2\pi = 1.99 \ \mathrm{THz}, \ \theta_1 = -0.9230$ and $\theta_2= 0.6188$.

The result of the optimization algorithm is limited by the minimal probe frequency $\omega_{\mathrm{pr}}$.
We have therefore examined the optimal protocol more closely for even smaller probe frequencies $\omega_{\mathrm{pr}} \in [0.0001 \omega_{\mathrm{pl}},~0.0005 \omega_{\mathrm{pl}}]$ and have reached a value of 134.9 for the effective Josephson coupling $J_{\mathrm{eff}}/J_0$ with the optimized driving parameters as we show in Fig.~\ref{fig:fig1}(a).
This probing regime corresponds to the left side of the shaded grey area, which indicates the probe frequencies explored in the optimization process.
\begin{figure*}
    \centering	
    \includegraphics[scale=1]{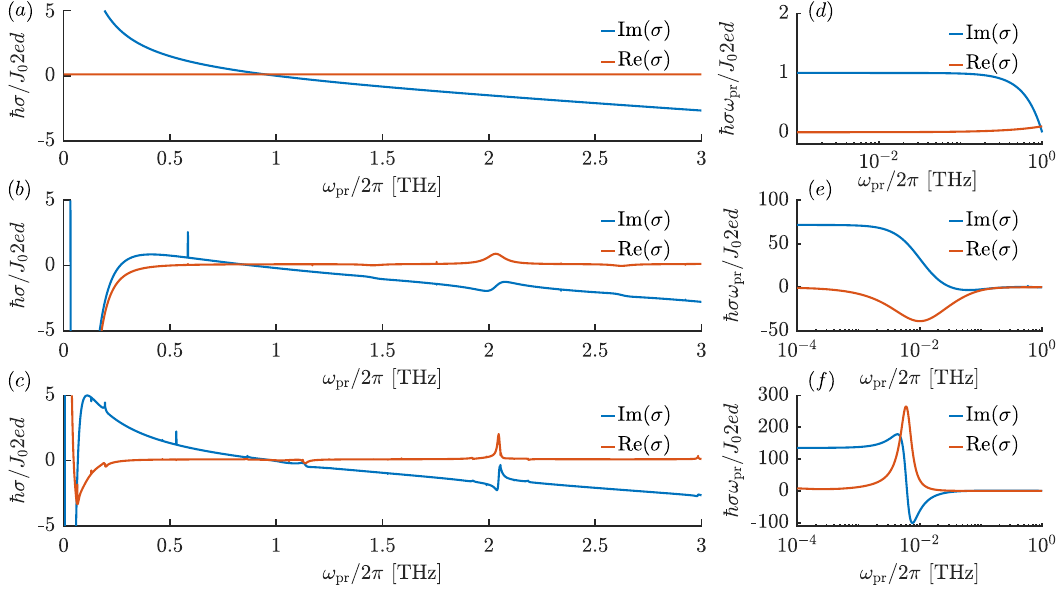}
    \caption{Numerical results for the real (red) and the imaginary (blue) part of the conductivity $\sigma(\omega_{\mathrm{pr}})$ for probe frequencies up to $3\times \omega_{\mathrm{pl}}/2\pi$. (a) Undriven case (b) optimal bichromatic driving protocol (c) optimal trichromatic driving protocol.
    In panels (d) - (f) we show the resolution of the low probe frequency limit $\omega_{\mathrm{pr}}$ $<\omega_{\mathrm{pl}} /2\pi$ for the real (red) and the imaginary (blue) part the product of the conductivity and $\omega_{\mathrm{pr}}$, i.e. of $\sigma(\omega_{\mathrm{pr}})\omega_{\mathrm{pr}}$. (d) Undriven case (e) optimal bichromatic driving protocol (f) optimal trichromatic driving protocol. Note that we rescale $\sigma$ with $\omega_{\mathrm{pr}}$ in order to resolve the $1/\omega_{\mathrm{pr}}$ divergence (d-f). }
    \label{fig:Conduc}
\end{figure*}
\section{Conductivity of optimally driven junctions}
In this section we investigate the frequency spectrum from $0$ to $3\times \omega_{\mathrm{pl}}$ and calculate numerically the real and imaginary part of the conductivity $\sigma(\omega_{\mathrm{pr}})$ according to Eq.~(\ref{eq:sig}). We show the results in Fig.~\ref{fig:Conduc} for the undriven case (a), the optimal bichromatic driving protocol and (c) the optimal trichromatic driving protocol. 
In the undriven case, see Fig.~\ref{fig:Conduc}(a), we obtain the expected results with a constant real part Re($\sigma$) $\equiv\sigma_1=\gamma$ and a $1/\omega_{\mathrm{pr}}$ divergence for the imaginary part Im($\sigma$) $\equiv \sigma_2$. In Fig.~\ref{fig:Conduc}(d) we display the product  $\sigma(\omega_{\mathrm{pr}})\omega_{\mathrm{pr}}$ in the low frequency limit on a logarithmic x-axis. We obtain a constant value  for the imaginary (blue) part, and a linear behavior for the real (red) part. 
We observe a peak around the probe frequency $\omega_{\mathrm{pr}}/2\pi \approx 2$ for both the real (red) and the imaginary (blue) part of the conductivity $\sigma$. This is near the second driving frequency $\omega_2$. Furthermore, while the imaginary part $\sigma_2$ has a zero crossing near $\omega_{\mathrm{pr}}/2\pi = 1$, similar to the undriven case, both $\sigma_1$ and $\sigma_2$ are negative for probe frequencies smaller than $0.2 \omega_{\mathrm{pl}}/2 \pi$. However, in contrast to the real part $\sigma_1$, the imaginary part $\sigma_2$ becomes positive again in the limit $\omega_{\mathrm{pr}}\rightarrow 0$, resulting in the large positive value of $J_{\mathrm{eff}}/J_0$ discussed before. In Fig.~\ref{fig:Conduc}(e) we resolve the low frequency limit of the product $\sigma(\omega_{\mathrm{pr}})\omega_{\mathrm{pr}}$ for an optimal bichromatic driving protocol with a logarithmic x-axis. We obtain a negative $\sigma_1$ which is related to optical gain with a maximal value of approximately $-30$ at $\omega_{\mathrm{pr}}/2 \pi \approx 0.02 $. In the low frequency limit of $\omega_{\mathrm{pr}}\rightarrow 0$, $\sigma_1$ convergences to 0, but it stays negative. The imaginary part $\sigma_2$ convergences to a value of approximately $70$ in the limit, which we defined as the effective Josephson coupling $J_{\mathrm{eff}}/J_0$ according to Eq.~(\ref{eq:jeff}).

In Fig.~\ref{fig:Conduc}(c) we show the conductivity of a junction driven by the optimal trichromatic protocol. Similar to the bichromatic protocol, the conductivity displays a peak near $\omega_{\mathrm{pr}}/2\pi \approx 2$. However, for the trichromatic protocol, the peak is narrower. Furthermore, the qualitative behavior of the imaginary part $\sigma_2$ is similar, with a negative regime below $\omega_{\mathrm{pr}}/2\pi \approx 0.1$, down to $\omega_{\mathrm{pr}}/2\pi \approx 0.01$. Below this frequency, $\sigma_2$ displays the $1/\omega_{\mathrm{pr}}$ dependence, with a large positive prefactor of around 135, discussed above. The real part of the conductivity $\sigma_1$ also changes sign, near $\omega_{\mathrm{pr}}/2\pi \approx 0.2$. However, in contrast to the bichromatic protocol, its sign changes again to a positive value near $\omega_{\mathrm{pr}}/2\pi \approx 0.01$, therefore not giving rise to optical gain.

\section{Conclusion}
\begin{figure}[!h]
    \centering	
    \includegraphics[scale=1]{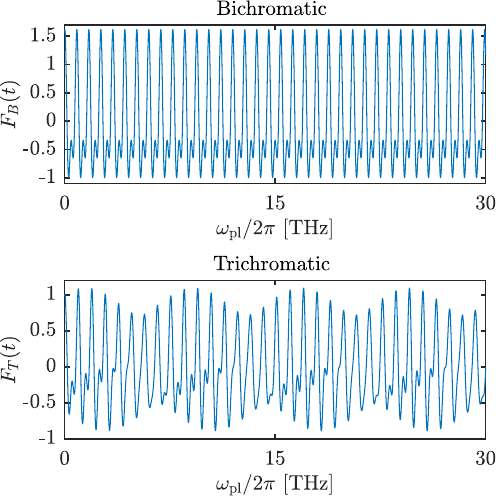}
    \caption{The optimized driving protocols $F_B(t)$ (top) and $F_T(t)$ (bottom).}
    \label{fig:driving vergleich}
\end{figure}

In conclusion, we have developed optimal control protocols to maximally enhance the DC transport across a Josephson junction. Specifically, we have identified bichromatic and trichromatic protocols, characterized by two or three frequencies, amplitudes, and phases, which modulate the Josephson coupling of the junction parametrically. To develop these protocols, we have utilized a Monte Carlo search, which optimizes the prefactor of the 1/$\omega_{\textrm{pr}}$ divergence of the imaginary part of the conductivity at low frequencies, i.e. the effective Josephson coupling. The two optimal protocols are depicted in Fig.~\ref{fig:driving vergleich}, as a visualization. 
The bichromatic protocols utilizes an driving frequency that is blue-detuned relative to the plasma frequency of the junction, and as a second frequency at twice the value of the base frequency. The trichromatic protocol also uses approximately  these two frequencies, with a third frequency that is not at an integer ratio to the base frequency, but rather near twice the plasma frequency. This trichromatic protocol shows a enhancement of the effective Josephson coupling of around 135, the bichromatic protocol shows an enhancement of around 70. This remarkable enhancement suggests that the phenomenon of light-enhanced superconductivity can be strongly advanced by utilizing multi-frequency protocols, as a next generation of dynamical control of materials via light.
\begin{acknowledgments}
This work is supported by the Deutsche Forschungsgemeinschaft (DFG) in the framework of SFB~925, Project No.~170620586, and the Cluster of Excellence ``Advanced Imaging of Matter" (EXC~2056), Project No.~390715994.
\end{acknowledgments}
\clearpage
\bibliography{references}

\appendix
\subsection{Analytical solution}\label{App:Analytical}
In this section we show the agreement of the numerical solution with the analytical solution in the equilibrium case. 
First we linearize Eq.~(\ref{Eq:DGL}) in the main text
\begin{equation}
    \Ddot{\varphi} + \gamma \Dot{\varphi} + \omega_{\mathrm{pl}}^2 \varphi= \omega_{\mathrm{pl}}^2 J(t) ,
\end{equation}
with $J(t)=J_0e^{-i\omega_{\mathrm{pr}}t}$.
We choose as our ansatz 
\begin{equation}
    \varphi(t)=\varphi_0e^{-i\omega_{\mathrm{pr}}t}
\end{equation}
Note that $\varphi_0 \propto \frac{E_0}{-i\omega_{\mathrm{pr}}}$ which implies that $J_0$ is proportional to $E_0$ in form of:
\begin{equation}
    E_0\left(\frac{-\omega^2_{\mathrm{pr}}-i\gamma \omega_{\mathrm{pr}}+\omega_{\mathrm{pl}^2}}{-i\omega_{\mathrm{pr}}}\right) \propto J_0
\end{equation}
with the conductivity $\sigma =\sigma_1+i\sigma_2$ as the proportionality constant.
We obtain the real part as $\sigma_1=\gamma$ and the imaginary part as 
\begin{equation}
\sigma_2=\left(\frac{\omega_{\mathrm{pl}}^2}{\omega_{\mathrm{pr}}} - \omega_{\mathrm{pr}} \right)
\label{eq:app-o2}
\end{equation}

\begin{figure}[h!]
    \centering
    \includegraphics{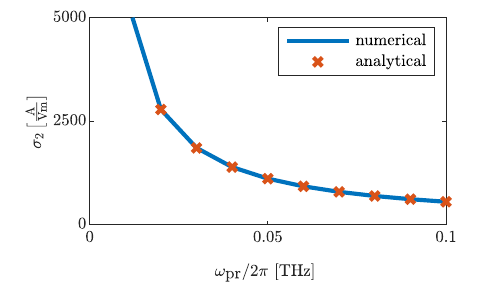}
    \caption{The imaginary conductivity of a single Josephson junction without driving. The numerical result matches the analytical solution Eq.~(\ref{eq:app-o2}).}
    \label{fig:Comp_a_n}
\end{figure}

In Fig.~\ref{fig:Comp_a_n} we show the imaginary part of the conductivity for both numerical and analytical solution which are in excellent agreement with each other. 

\subsection{Floquet analysis}\label{A2. Floquet}

We briefly explain the Floquet analysis~\cite{floquet1,floquet2} that we use to identify protocols that generate strong heating.
The Floquet theorem enables statements about homogeneous, ordinary differential equation systems of the form
\begin{equation}
     \ddot{\Vec{x}}(t) = \mathbf{A}(t) \Vec{x}(t)
\end{equation}
with a periodic coefficient matrix $\mathbf{A}(t)=\mathbf{A}(t+T)$. The fundamental matrix takes the following form:
\begin{equation}
   \mathbf{\Phi} (t) = P(t) e^{t R} \label{eq:floquet}
\end{equation}
with $P(t)=P(t+T)$ a time periodic function. The eigenvalues $\lambda_i$ of $R$ are the Floquet-exponents.
If the absolute value of $|\lambda_i|$ is larger than 1 this indicates dynamical instability. This leads to a  diverging Eq.~(\ref{eq:floquet}).
In our model we linearize Eq.~(\ref{Eq:DGL}) and set $J=0$, add parametric driving to the dynamics. We numerically solve the equation of motion
\begin{equation}
    \left(  \begin{array}{cc} 
    \dot{\varphi_1} & \dot{\varphi_2}\\ 
    \ddot{\varphi_1} & \ddot{\varphi_2}   
\end{array}\right)
= \left(\begin{array}{cc} 
    0 & 1\\ 
    -\omega_{pl}^2 \left[1+\text{A}\cos(\omega t)\right] & -\gamma 
\end{array} \right)
\left(  \begin{array}{cc} 
    \varphi_1 & \varphi_2\\ 
    \dot{\varphi_1} & \dot{\varphi_2} 
\end{array}\right)
\end{equation}
over a period from $t=0$ to $t=2\pi / \omega$ with the initial condition \begin{equation}
    \left( \varphi_1(0),\dot{ \varphi_1}(0)\right)=\left(1,0\right)
\end{equation}
\begin{equation}  
   \left(\varphi_2(0),\dot{ \varphi_2}(0)\right)=\left(0,1\right)
\end{equation}
We obtain the eigenvalues of the resulting matrix 
\begin{equation}
    \lambda_{1,2}=\pm 
    \sqrt{\frac{1}{4}\left(\varphi_1-\dot{\varphi}_2\right)^2+ \dot{\varphi}_1\varphi_2 }+\frac{1}{2}\varphi_1+\dot{\varphi}_2
\end{equation}
which are the Floquet exponents.

\subsection{Impact of the initial state}\label{Initial}
In this section, we discuss the impact of the initial state. For a bichromatic driving protocol, the choice of the initial state does not affect the ability of the algorithm to find an optimal protocol.
\begin{figure}[h!]
    \centering
    \includegraphics[scale=1]{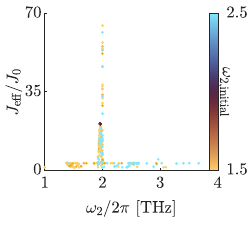}\llap{\parbox[b]{8cm}{\textcolor{black}{(a)}\\\rule{0ex}{3.8cm}}} \hspace{0.05cm} \includegraphics[scale=1]{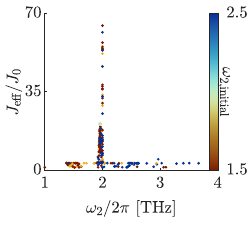}\llap{\parbox[b]{8cm}{\textcolor{black}{(b)}\\\rule{0ex}{3.8cm}}}
	\caption{$\omega_{2,\mathrm{initial}}$ as a function of the final $\omega_2$ and the corresponding effective Josephson coupling $J_{\mathrm{eff}}/J_0$ for 600 optimization trajectories with a bichromatic driving protocol $F_B(t)$.}
    \label{fig:bi_start}
\end{figure}
Our choice of the initial driving parameters $q_0$ are based on previous results~\cite{Okamoto-2016,Okamoto-2017}.
We examine 11 different initial parameter sets $q_0$. We choose as combinations for $(\omega_1/\omega_{\mathrm{pl}},\omega_2/\omega_{\mathrm{pl}})=\{(0.7,1.7)$, $(0.7,2.5)$, $(1.15,1.5)$, $(1.15,1.7)$ and $(1.15,2.5)\}$ for two driving amplitudes $A_1 =0.1$ and $A_1=0.6$

\begin{table}[h!]
    \centering
    \begin{tabular}{|c|c|c|c|c|}
    \hline
        $A_1$ & $A_2$ & $\omega_{1} [\mathrm{THz}/2\pi]$& $\omega_{2}[\mathrm{THz}/2\pi]$ & $\theta$  \\
        \hline
         0.1 & 0.1 & 0.7 & 1.7 & 0  \\
          \hline
         0.6 & 0.1 & 0.7 & 1.7 & 0  \\
          \hline
         0.1 & 0.1 & 0.7 & 2.5 & 0  \\
          \hline
         0.6 & 0.1 & 0.7 & 2.5 & 0 \\
          \hline
         0.1 & 0.1 & 1.15 & 1.5 & 0  \\
          \hline
         0.6 & 0.1 & 1.15 & 1.5 & 0  \\
          \hline
         0.1 & 0.1 & 1.15 & 1.7 & 0  \\
         \hline
         0.6 & 0.1 & 1.15 & 1.7 & 0  \\
          \hline
         0.1 & 0.1 & 1.15 & 2.5 & 0  \\
         \hline
         0.6 & 0.1 & 1.15 & 2.5 & 0  \\
         \hline
         0.71 & 0.25 & 1.05 & 2.06 & -3.26  \\
         \hline
    \end{tabular}
    \caption{Initial driving parameter $q_0$ used for the optimization of a bichromatic driving protocol $F_B(t)$.}
    \label{tab:bichrom}
\end{table}

In every initial set $A_2$ was set to 0.1 and we tested two different values for $A_1=(0.1,0.6)$. The phase difference $\theta$ was chosen to be 0.

In Fig.~\ref{fig:bi_start} we present Fig.~\ref{fig:fig3}(a) once again, as previously presented in the main text. We show $\omega_{2,\mathrm{initial}}$ as a function of the final $\omega_2$ and the corresponding effective Josephson coupling $J_{\mathrm{eff}}/J_0$ in the optimization.
Regardless of the choice of the initial $\omega_2$, the parameter space was screened and a significant increase in coupling was found around the second harmonic.

In contrast, the initial state, more precisely the value of $\omega_3$, in the trichromatic driving approach has a significant influence on the success of the optimization algorithm.
We examine the following 13 different initial parameter sets:

\begin{table}[h!]
    \centering
\begin{tabular}{|c|c|c|c|c|c|c|c|}
    \hline
        $A_1$ & $A_2$ & $A_3$ & $\omega_{1}  [\mathrm{THz}/2\pi]$& $\omega_{2}[\mathrm{THz}/2\pi]$ & $\omega_3[\mathrm{THz}/2\pi]$ &$\theta_1$ & $\theta_2$ \\
    \hline
    0.4 & 0.2 & 0.1 & 1.15 & 2.3 & 0.7 & 0 & 0 \\
    \hline
    0.4 & 0.2 & 0.1 & 1.15 & 2.3 & 1.15 & 0 & 0 \\
    \hline
    0.4 & 0.2 & 0.1 & 1.15 & 2.3 & 1.3 & 0 & 0 \\
    \hline
    0.4 & 0.2 & 0.1 & 1.15 & 2.3 & 1.7 & 0 & 0 \\
    \hline
    0.4 & 0.2 & 0.1 & 1.15 & 2.3 & 2.3 & 0 & 0 \\
    \hline
    0.4 & 0.2 & 0.1 & 1.15 & 2.3 & 2.7 & 0 & 0 \\
    \hline
    0.4 & 0.2 & 0.1 & 1.15 & 2.3 & 3.3 & 0 & 0 \\
    \hline
    0.4 & 0.2 & 0.1 & 1.15 & 2.3 & 3.7 & 0 & 0 \\
    \hline
    0.4 & 0.2 & 0.1 & 1.15 & 2.3 & 4.3 & 0 & 0 \\
    \hline
    0.67 & 0.46 & 0.36 & 1.12 & 2.24 & 3.8 & -0.11 & 1.95 \\
    \hline
    0.57 & 0.28 & 0.19 & 1.05 & 2.1 & 1.98 & -0.89 & 0.95 \\
    \hline
    0.61 & 0.27 & 0.16 & 1.06 & 2.12 & 2.01 & -0.95 & 0.64 \\
    \hline
\end{tabular}
\caption{Initial driving parameter $q_0$ used for the optimization of a trichromatic driving protocol $F_T(t)$.}
\label{tab:trichrom}
\end{table}

\begin{figure}[h!]
    \centering
    \includegraphics[scale=1]{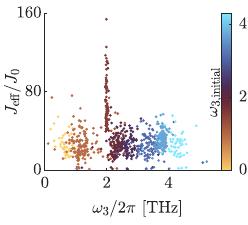} 
	\caption{$\omega_{3,\mathrm{initial}}$ as a function of the final $\omega_3$ and the corresponding effective Josephson coupling $J_{\mathrm{eff}}/J_0$ for 950 optimization trajectories with a trichromatic driving protocol $F_T(t)$.}
    \label{fig:thri_start}
\end{figure}

We illustrate in Fig.~\ref{fig:thri_start} $\omega_{3,\mathrm{initial}}$ as a function of the final $\omega_3$ and the corresponding effective Josephson coupling $J_{\mathrm{eff}}/J_0$ in the optimization. We observe that the choice of the initial $\omega_3$ has an influence on the range in which the parameter space of $\omega_3$ is screened. The operating distance is significantly lower. We suspect that this could be due to the restrictive conditions on the parameters $q$ such as the limited driving protocol $F(t)>-1$ or our Floquet analysis to avoid dynamical instability which makes the hopping over forbidden areas more difficult or impossible.

\end{document}